\documentclass[]{aa}  
\usepackage{graphicx}
\usepackage[varg]{txfonts}
\usepackage{xspace}
\usepackage{color}
\definecolor{DarkGreen}{rgb}{0.0,0.4,0.0}  

\newcommand{\kms}{~km~s$^{-1}$\xspace} 
\makeatletter
\newcommand*{\rom}[1]{\expandafter\@slowromancap\romannumeral #1@}
\makeatother

\begin{document}

\title{Erupting Magnetic Flux Rope Affects Running Penumbral Waves}

\author{Wensi Wang\inst{1}
	\and 
	Rui Liu\inst{1,2,3}
	}

\institute{CAS Key Laboratory of Geospace Environment, Department of Geophysics and Planetary Sciences\\
University of Science and Technology of China, Hefei, 230026, China\\ \email{rliu@ustc.edu.cn}
\and
CAS Center for Excellence in Comparative Planetology, Hefei 230026, China
\and
Mengcheng National Geophysical Observatory\\
University of Science and Technology of China, Mengcheng 233500, China
}

\date{Received Oct 21, 2020; accepted Jan 12, 2021}
\abstract
{It is well known that solar flares have broad impacts on the low atmosphere, but it is largely unknown how they affect sunspot waves and oscillations. It is also under debate as to whether the flare-induced photospheric changes are due to the momentum conservation with coronal mass ejections or due to magnetic reconnection.} 
{To shed light on the so-called ``back reaction'' of solar eruptions, we investigated how running penumbral waves (RPWs) at one foot of an erupting magnetic flux rope (MFR) responds to the rope buildup and subsequent erosion.} 
{We used UV/EUV images from the Atmospheric Imaging Assembly on-board the Solar Dynamics Observatory (SDO) to explore the changing behaviors of RPWs in response to the MFR evolution, and 135-s vector magnetograms from the SDO Helioseismic and Magnetic Imager to analyze the changes in photospheric magnetic field during the eruption.}
{During the rope-buildup stage, the western foot of the rope, which is completely enclosed by a hooked ribbon, expands rapidly and consequently overlaps a sunspot penumbra. This converts the original penumbral field into the rope field, which is associated with a transient increase in electric currents flowing through the ribbon-swept penumbral region. During the rope-erosion stage, the rope foot shrinks as the eastern section of the hooked ribbon slowly sweeps the same penumbral region, where the rope field is converted into flare loops. This conversion induces mixed effects on the photospheric field inclination but heats up the low atmosphere at the footpoints of these flare loops to transition-region temperatures, therefore resulting in the post-eruption RPWs with an enhanced contrast in the 1600~{\AA} passband and an extended bandwidth to low frequencies at 3--5~mHz, compared with the pre-eruption RPWs that peak at 6~mHz.} 
{This observation clearly demonstrates that it is the magnetic reconnection in the corona that impacts the low atmosphere and leads to the changing behaviors of RPWs, which, in turn, offer a new window to diagnose flare reconnections. }

\keywords{sunspot waves -- solar eruptions -- magnetic reconnection}

\maketitle
	
\section{Introduction} \label{sec:intro}
Coronal mass ejections (CMEs) are a sudden, explosive release of magnetic stress building up in the corona through the gradual evolution of photospheric magnetic field, as driven by turbulent motions in photospheric and sub-photospheric layers \citep{Forbes2006}. The core of CMEs is often demonstrated to possess a twisted structure known as a magnetic flux rope \citep[MFR;][]{Liu2020}. Since the tenuous corona is over six orders of magnitude less dense than the photosphere, it is as hard to imagine that a CME could inflict any appreciable impact on the photosphere as ``the tail wags the dog'' \citep{Aunlanier2016}. 

Surprisingly, coronal eruptions are indeed associated with rapid changes in the photosphere. For example, step-wise, permanent field changes in photospheric field are found in numerous flares \citep[see the review by][]{Toriumi&Wang2019}. These changes are often manifested in white-light structures of sunspots, with enhanced (i.e., darkened) penumbral structures near the flaring polarity inversion lines (PILs) and decayed penumbral structure in the periphery of $\delta$-spots \citep[e.g.,][]{liu2005rapid}. The former is related to magnetic field turning more horizontal, with the magnitude of horizontal field component increasing by a few hundred Gauss \citep[e.g.,][]{Sun2012}; the latter is related to magnetic field turning more vertical \citep[e.g.,][]{WangH2012}. 

Near the flaring PIL, that the magnetic field turns more horizontal is envisaged to be the downward collapsing of magnetic loops, which is suggested to be caused either by the momentum conservation associated with CMEs \citep{Hudson2008,Fisher2012} or by a decrease in magnetic pressure associated with the flaring release of free magnetic energy \citep{Hudson2000}. With a 3D magnetohydrodynamics simulation, however, \citet{Barczynski2019} argued that photospheric horizontal fields are enhanced by reconnection-driven contraction of sheared flare loops, rather than by a downward Lorentz force as a result of momentum conservation, because the true Lorentz force density derived in a 3D magnetic field could differ significantly from that obtained from photospheric vector magnetograms as a proxy.

Much attention has been given to photospheric regions swept by flare ribbons, which highlight the footpoints of newly reconnected magnetic field lines. \citet{WangH2012} demonstrated that in the peripheral penumbrae that are swept by flare ribbons, dark fibrils completely disappear while bright grains evolve into faculae where the magnetic field becomes more vertical. \cite{Bi2016} reported that the a sunspot suddenly reverse its rotational direction during a flare. \citet{LiuC2016} found that a solar flare induces differential rotation of a sunspot as one flare ribbon sweeps the sunspot. It is postulated that the rotation is driven by a torque arising with the flare-related restructuring of coronal magnetic field. In the same event, \citet{liu2018evolution} found that changes of photospheric fields occur nearly instantaneously at the arrival of the flare ribbon front. In the ribbon-swept sunspot region, photospheric field turns temporarily more vertical, associated with a sudden counterclockwise rotation of field vectors by 12--20$^\circ$ in azimuth degree \citep[see also][]{Xu2018}, and then either recovers or turns more horizontal than before the ribbon arrival.

Occasionally, solar flares are found to excite waves in the quiet Sun from the interior up to coronal heights. \citet{Kosovichev&Zharkova1998} found that an X2.6-class flare impacts the Sun's interior by generating seismic waves similar to earthquakes, known as `sunquakes', whose generation mechanisms are still under debate. In the recent X9.3-class flare, the locations of a few sunquakes and H$\alpha$ kernels are believed to be associated with the footpoints of MFRs formed before the flare onset \citep{Zharkov2020}. \citet{Kosovichev&Sekii2007} reported that an X3.4-class two-ribbon flare excited oscillations in the chromosphere of a nearby sunspot. The oscillations have larger amplitudes and higher frequencies than before the flare, and appear to propagate through the umbra in the direction away from flare ribbons. 

Meanwhile, various waves and oscillations are constantly present in sunspots during the quiet period \citep[see the review by][]{Khomenko&Collados2015}. These are believed to be excited by convection in the sub-photospheric layers, but modulated by strong magnetic fields in sunspots. In particular, running penumbral waves (RPWs) appear to emit from the umbra as propagating brightness disturbances that expand concentrically at a typical speed of 10--15~\kms and extend slightly beyond the outer penumbral boundary. These are interpreted as slow magneto-acoustic waves propagating longitudinally along magnetic field lines, with the wavefront expanding along the inclined field in the penumbra. In the H$\alpha$ line core, RPWs have a dominating period of five minutes in the inner penumbra, but the period increases toward the outer penumbra \citep{jess2013influence}. This is because only those with frequencies above the acoustic cutoff  $f_c=g\gamma/\sqrt{2\pi C_s}$ can propagate upward in the presence of gravitation. Here $C_s=\sqrt{\gamma R T/\mu}$ is the acoustic speed, in which $\mu$ is the mean molecular weight. The effective gravity $g$ on a particular magnetic field line is modulated by the field inclination angle $\theta$ with respect to the solar normal, i.e., $g=g_0\,\cos\theta$, where $g_0$ is gravitational acceleration \citep{bel1977analytical}. Thus, the cut-off frequency is reduced by the increasing field inclination towards the outer penumbra \citep{jess2013influence}.

However, it is a largely uncharted area in regard to how the flare-ribbon sweeping affects sunspot waves and oscillations. The M3.7-class flare on 2015 November 4 provides such an opportunity to better understand how the low atmosphere responds to coronal eruptions, as one of the erupting MFR's feet, which are formed during the eruption and outlined by closed, hooked ribbons \citep[][hereafter Paper I]{wang2017buildup}, temporarily overlaps part of a sunspot penumbra. In the sections that follow, we present the observations of how the evolution of the MFR affects both the propagation and spectrum of RPWs (\S~\ref{sec:obs}) and then discuss the results and their implications (\S~\ref{sec:dis}).

\section{Observations \& Analyses} \label{sec:obs}
\subsection{Instruments and Data}
The MFR eruption as investigated in detail in Paper I is mainly observed in UV/EUV by the Atmospheric Imaging Assembly \citep[AIA;][]{Lemen2012} onboard the Solar Dynamics Observatory \citep[SDO;][]{Pesnell2012}. AIA takes full-disk images with a spatial scale of $0''.6$ pixel$^{-1}$ and a cadence of 12 s for EUV and 24s for UV passbands. Among its 7 EUV and 2 UV passbands, we fixated our attention on 131, 171, and 1600~{\AA} in this study. The 131~{\AA} passband is dominated by Fe~\rom{21} ($\log T = 7.05$) in flares, but also has a significant contribution from Fe~\rom{8} ($\log T = 5.6$) \citep{ODwyer2010}. The 171~{\AA} passband is dominated by Fe~\rom{9} ($\log T = 5.8$). Like Fe~\rom{8}, Fe~\rom{9} features quiet corona and upper transition region. The 1600~{\AA} passband is dominated by the C~\rom{4} 1550~{\AA} doublet ($\log T = 5.0$) and Si~\rom{1} continua \citep{Simoes2019}, which feature the transition region and upper photosphere, respectively.

The changes of photospheric magnetic field during the flare are investigated with vector magnetograms at 135-s cadence obtained by the Helioseismic and Magnetic Imager \citep[HMI;][]{Scherrer2012} onboard SDO. 90 high-cadence vector magnetograms are available during the time interval from 12:00 UT to 15:30 UT. Although the high-cadence magnetograms are subject to higher uncertainties than the conventional ones at 720-s cadence, they agree well with each other in strong-field regions \citep{sun2017investigating}. We hence focused on the sunspot in this study. The SSW procedure \texttt{hmi\_disambig.pro} is applied to solve the 180~deg ambiguity, and SSW procedures \texttt{hmi\_b2ptr.pro} and \texttt{bvec\_errorprop.pro} are applied to obtain the heliographic components of vector field and corresponding errors. We then used these errors to constrain a Monte Carlo simulation of electric current density $J_r$. Randomly varying vector field at each pixel by $10^5$ times, we ended up with a Gaussian distribution of $J_r$ at each pixel, whose mean and standard deviation are then taken as the $J_r$ value and error, respectively.

\subsection{MFR buildup and erosion during the 2015 November 4 eruption}
The dynamic formation of the MFR on 2015 November 4 is associated with a long-duration flare and a halo CME. Passing through spacecrafts near Earth three days later, the interplanetary CME manifests itself as a typical magnetic cloud (Figure 5 in Paper I), which possesses a similar twist profile as derived from its buildup process in the low corona (Figure 4 in Paper I) and whose axis is oriented similarly as the connection of the MFR's two feet on the Sun (Supplementary Figure 7 in Paper I). 

Compared with double J-shaped ribbons with open hooks \citep[e.g.,][]{Janvier2014,Zhao2016}, the 2015 November 4 event features a new flare-ribbon morphology --- double J-shaped ribbons with two closed hooks \citep{wang2017buildup,Li2017,Zemanova2019}, labeled as `FP+' and `FP-' in Figure~\ref{fig:ribbon-morphology}a. Initially the two flare ribbons extend along, but not move away from, the PIL. When these ribbons, termed `straight ribbons' hereafter, extend to their full lengths, a closed hook (termed `hooked ribbon' hereafter) expands from a point-like brightening at the far end of each straight ribbon. The expansion is associated with the development of coronal dimming inside each hook, the rapid separation of two straight ribbons, and impulsive HXR bursts (Figures 2 and 3 in Paper I). The conjugated pair of dimmings are clearly visible and fully enclosed by the brightened hooks in all of AIA's EUV passbands, indicating mass depletion along the MFR legs. That each closed hook develops from a point-like brightening strongly suggests that the bulk of the MFR is formed on the fly. Counting magnetic flux through the MFR's feet as enclosed by the hooked ribbons and flux through the area swept by straight ribbons, Paper I reveals that the MFR has a highly twisted core (up to $\sim\,$10 turns) and less twisted ( $\sim\,$4 turns) outer shells, which is consistent with the prediction of theoretical and modeling studies that open and closed hooks are associated with moderate ($\sim\,$1 turn) and strong ($\ge2$ turns) twist of the flux rope, respectively \citep{Demoulin1996,Pariat&Demoulin2012,Zhao2016}.

In this paper, we focus on the MFR's western foot FP-, which is identified by combining the hooked ribbon in 1600~{\AA} and the coronal dimming in 304 and 335~{\AA} (see Methods in Paper I).  When it is full-fledged, FP- overlaps the penumbra of a sunspot of negative polarity in the NOAA active region 12443 (Figure~\ref{fig:ribbon-morphology}a). According to the temporal variation of the magnetic flux through the foot, $\Phi_\mathrm{FP-}$, the MFR evolves in two stages: a buildup stage until 13:52 UT during which the foot expands and $\Phi_\mathrm{FP-}$ increases monotonically, followed by an erosion stage during which the foot shrinks and $\Phi_\mathrm{FP-}$ decreases (Figure~\ref{fig:fp-evolution}a, see also Supplementary Figure 4 in Paper I). 

Associated with the expansion of FP-, the southern section of the hooked ribbon moves southeastward into the penumbra during 13:40--13:52 UT (Figure~\ref{fig:fp-evolution}(c--d), denoted as `HR$_\mathrm{S}$' hereafter). HR$_\mathrm{S}$ reaches the inner pneumbra boundary, as outlined by the red dashed curve in Figure~\ref{fig:fp-evolution}d. Bounded by the outer penumbral boundary, the ribbon-swept penumbral region is outline by the white closed curve in Figure~\ref{fig:fp-evolution}g, termed penumbral region of interest (PROI) hereafter. The hooked ribbon dims from about 13:45 UT onward. But at about 14:03 UT, the eastern section of the hooked ribbon brightens up again and moves southwestward (marked by a red dotted line in Figure~\ref{fig:fp-evolution}(e--g) and denoted as `HR$_\mathrm{E}$' hereafter), sweeping the PROI for a second time until 14:50 UT. In the wake of this new ribbon sweeping, the PROI emits diffusively (Figure~\ref{fig:fp-evolution}g). The average 1600~{\AA} brightness in the PROI (blue curve in Figure~\ref{fig:fp-evolution}a) shows a spiky peak associated with the sweeping by HR$_\mathrm{S}$, and then a smaller peak with a gradual rise and a long decay, associated with the sweeping by HR$_\mathrm{E}$.  

Figure~\ref{fig:connectivity} demonstrates that during the rope-buildup stage, the hooked ribbons highlight the footpoints of the eruptive structure (Figure~\ref{fig:connectivity}(a \& c); see also Figure 1 in Paper I); and that during the rope-erosion stage, the region swept by HR$_\mathrm{E}$ has become the footpoints of post-flare loops that are connected to the straight ribbon of positive polarity (Figure~\ref{fig:connectivity}(d--f)). There are also post-flare loops connecting the straight ribbon of positive polarity with that of negative polarity, as well as loops connecting the hooked ribbon of FP+ in the east with the straight ribbon of negative polarity. Such connectivities are further clarified when these post-flare loops, somewhat diffuse in 131 and 94~{\AA} (Figure~\ref{fig:connectivity}(d \& e)), later cool down and appear more clearly defined in 171~{\AA} (Figure~\ref{fig:connectivity}f).

\subsection{Running Penumbra Waves}

\subsubsection{RPWs before eruption} \label{subsec:pre-erupt-rpw}
The sunspot of interest has a round shape overall, but the umbra has a mushroom-shaped `offshoot' extruding into the penumbra in the northeast quadrant, which breaks the circular symmetry. RPWs can be seen in the form of arc-shaped bright fronts propagating radially outwards from the inner to the outer penumbral boundary. These fronts are visible in running difference images in almost all AIA's UV/EUV passbands (Figure~\ref{fig:ribbon-morphology} and accompanying animation), except for the 94~{\AA} passband, which has poor signal-to-noise ratio and mostly features hot coronal plasmas of about 6~MK. The 171~{\AA} passband shows the fronts in best contrast. The disturbances appear to propagate outwards in all directions in the photosphere and chromosphere, as demonstrated in the 1600 and 304~{\AA} passbands; but in the higher atmosphere, e.g., in 171~{\AA}, the arc-shaped fronts seem to avoid the umbral offshoot in the northeast quadrant, only propagating in the other three quadrants. 

We adopted a virtual circular slit running through the penumbra (Figure~\ref{fig:wave-quiet}b) to further quantify the propagation of RPWs. Along the slit, azimuth degree increases clockwise from 0 deg in the north. Applying the circular slit to pre-eruption running difference 171~{\AA} images during 12:00--13:30 UT, we obtained a stack plot as shown in Figure~\ref{fig:wave-quiet}d. This time-distance map clearly shows that RPWs propagate roughly within an azimuth range of [0--240] deg. Applying the same circular slit to pre-eruption HMI vector magnetograms, we obtained time-distance maps of field inclination angle $\theta$ (Figure~\ref{fig:wave-quiet}d) and horizontal component $B_h$ (Figure~\ref{fig:wave-quiet}e). $\theta$ is relative to the solar normal, ranging between 0 and 180 deg. At 0 (180) deg magnetic field vectors are directed vertically upward (downward), and at 90 deg, vectors are parallel to the surface. It is clear that magnetic field vectors are highly inclined toward the surface where RPWs are observed in 171~{\AA}. The umbral offshoot, where $\mathbf{B}$ is almost vertical, appears to be the reason that RPWs are not visible in this direction.

\subsubsection{RPWs during eruption} \label{subsec:post-erupt-rpw}
To investigate the influence of the MFR eruption on RPWs, We applied a virtual linear slit across the PROI (Figure~\ref{fig:fp-evolution}d) to AIA 1600, 304, and 171~{\AA} running difference images. Along the slit, the PROI is located between about 4--9 Mm. In the resultant time-distance maps, the two episodes of ribbon sweeping through the PROI are shown as two white stripes at about 13:40 and 14:30 UT, respectively (Figure~\ref{fig:wave-erupt}(d--g)). 

\begin{figure*}
	\centering
	\includegraphics[width=\textwidth]{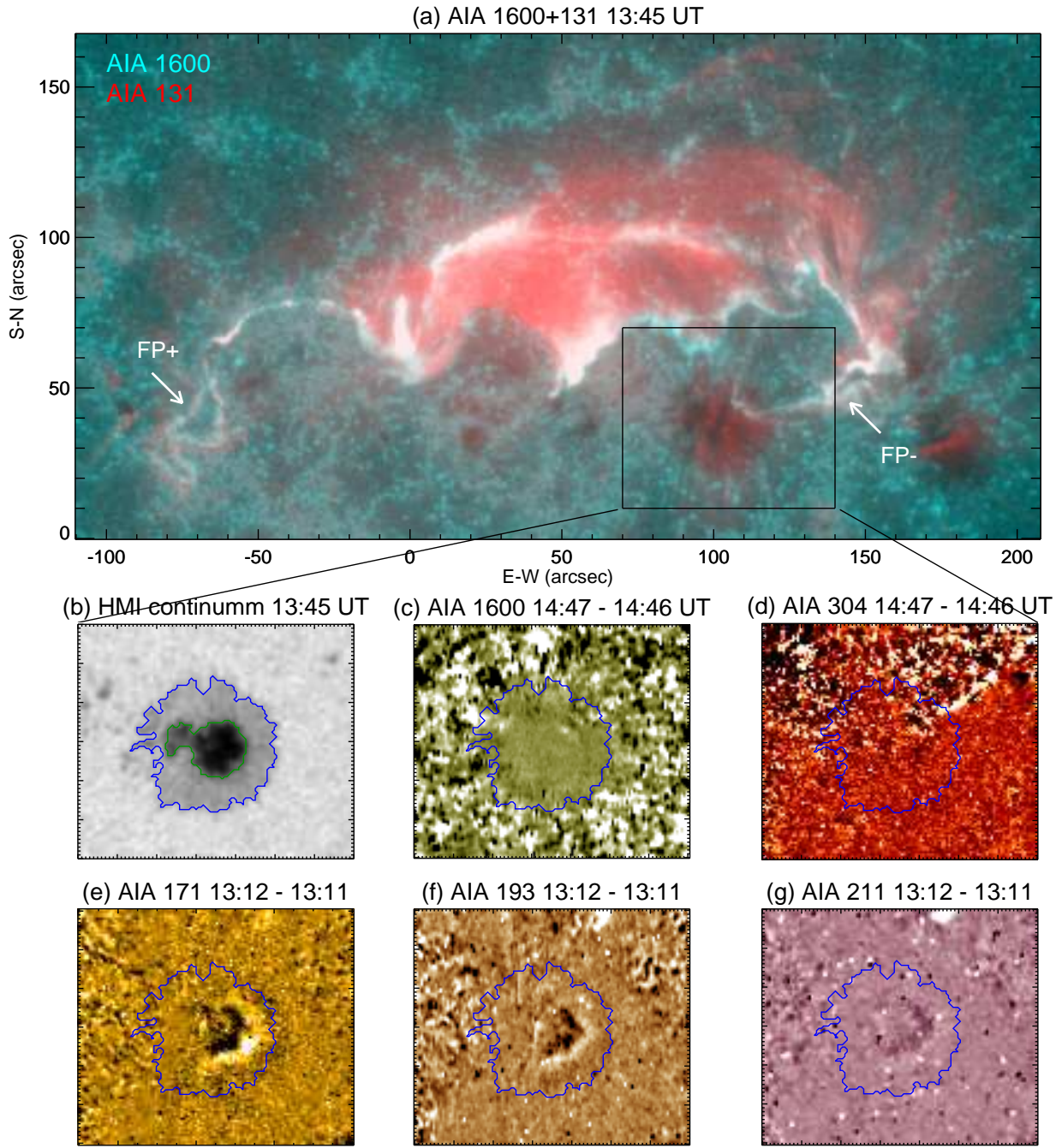}
	\caption{Flare ribbons in relation to the sunspot of interest. Panel (a) combines a 1600~{\AA} (cyan) and a 131~{\AA} (red) image to show the flare-ribbon morphology with two closed hooks attached to the far ends of two main ribbons. The MFR foot associated with negative (positive) polarity is labeled as FP- (FP+). Panels (b--g) zoom into the sunspot, with the field of view indicated by the rectangle in (a). Panel (b) presents the sunspot observed by the HMI continuum, with the green (blue) contour marking the inner (outer) penumbral boundary. Panels (c--g) show the RPW fronts in difference images from various AIA passbands. Consisting of snapshots in the same format as the figure, the online animation shows the whole flaring process and the RPWs in the sunspot, covering the time period from 12:00 to 16:00 UT. \label{fig:ribbon-morphology}}
\end{figure*}

\begin{figure*}
  \centering
  \includegraphics[width=\textwidth]{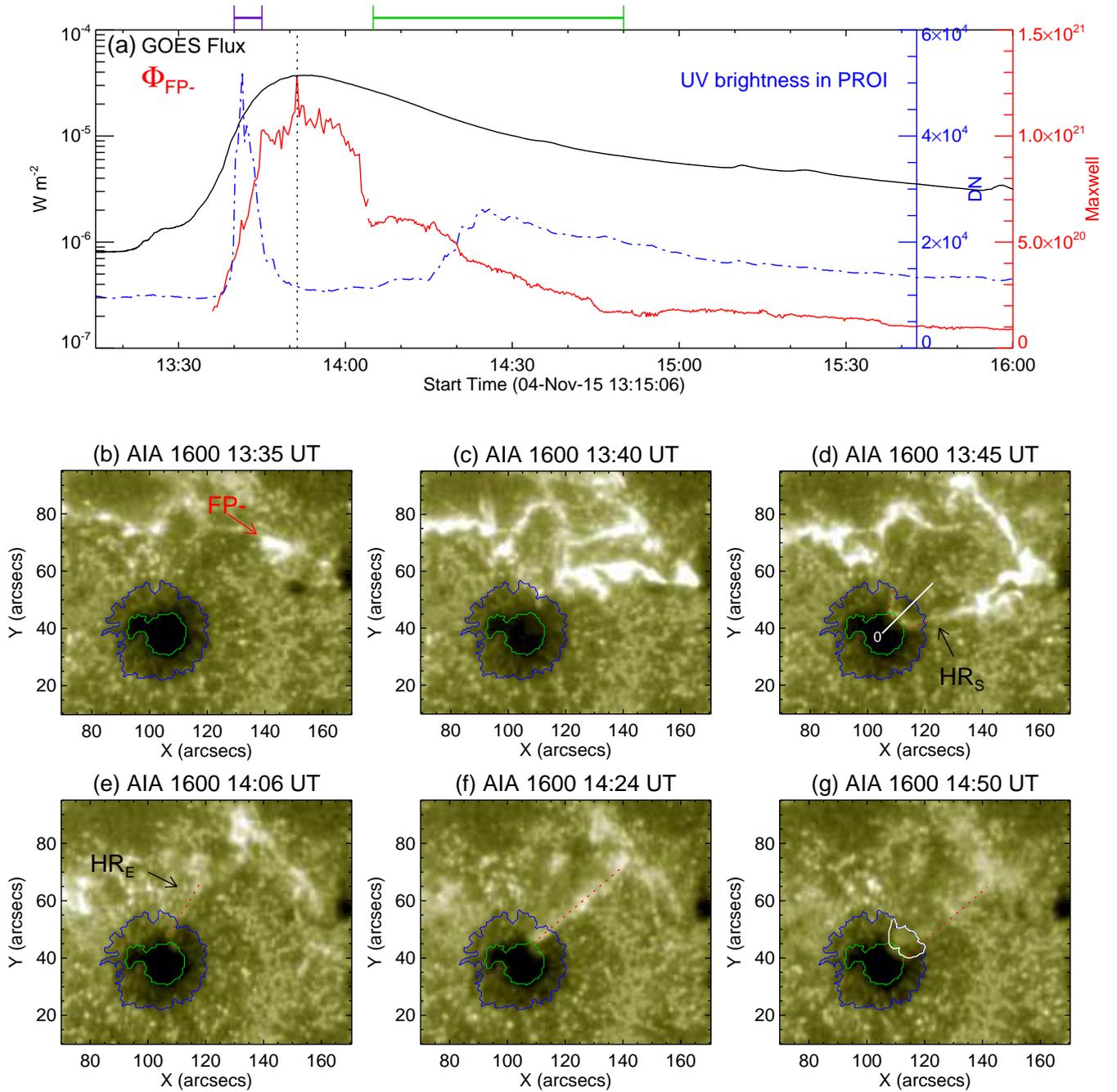}
  \caption{Evolution of the hooked ribbon enclosing the MFR foot FP-. Panel (a) shows lightcurves of GOES 1--8~{\AA} flux (black; scaled by the left $y$-axis), magnetic flux through the MFR foot (red; scaled by the red $y$-axis on the right), and the mean brightness of the PROI in 1600~{\AA} (blue; scaled by the blue $y$-axis on the right ). The purple (green) bar on the top indicates the time interval during which the hooked ribbon sweeps through the PROI during the rope-buildup (erosion) stage. Panels (b--g) show snapshots of AIA 1600~{\AA} images. The green (blue) contour marks the inner (outer) penumbral boundary. Red dashed lines outline sections of the hooked ribbon that move into the penumbra. The PROI as outlined by the red dashed line in (d) and the outer penumbral boundary is indicated by the white closed curve in (g). The white line in (d) indicates a virtual linear slit adopted to make time-distance maps in Figures~\ref{fig:wave-erupt} and \ref{fig:deltab}, with `0' labeling its starting point. The online animation shows AIA 1700, 1600, and 304~{\AA} images during 13:00--16:00 UT, with panel (a) severing as a timeline reference.  \label{fig:fp-evolution}} 
\end{figure*}

\begin{figure*}
	\centering
    \includegraphics[width=\textwidth]{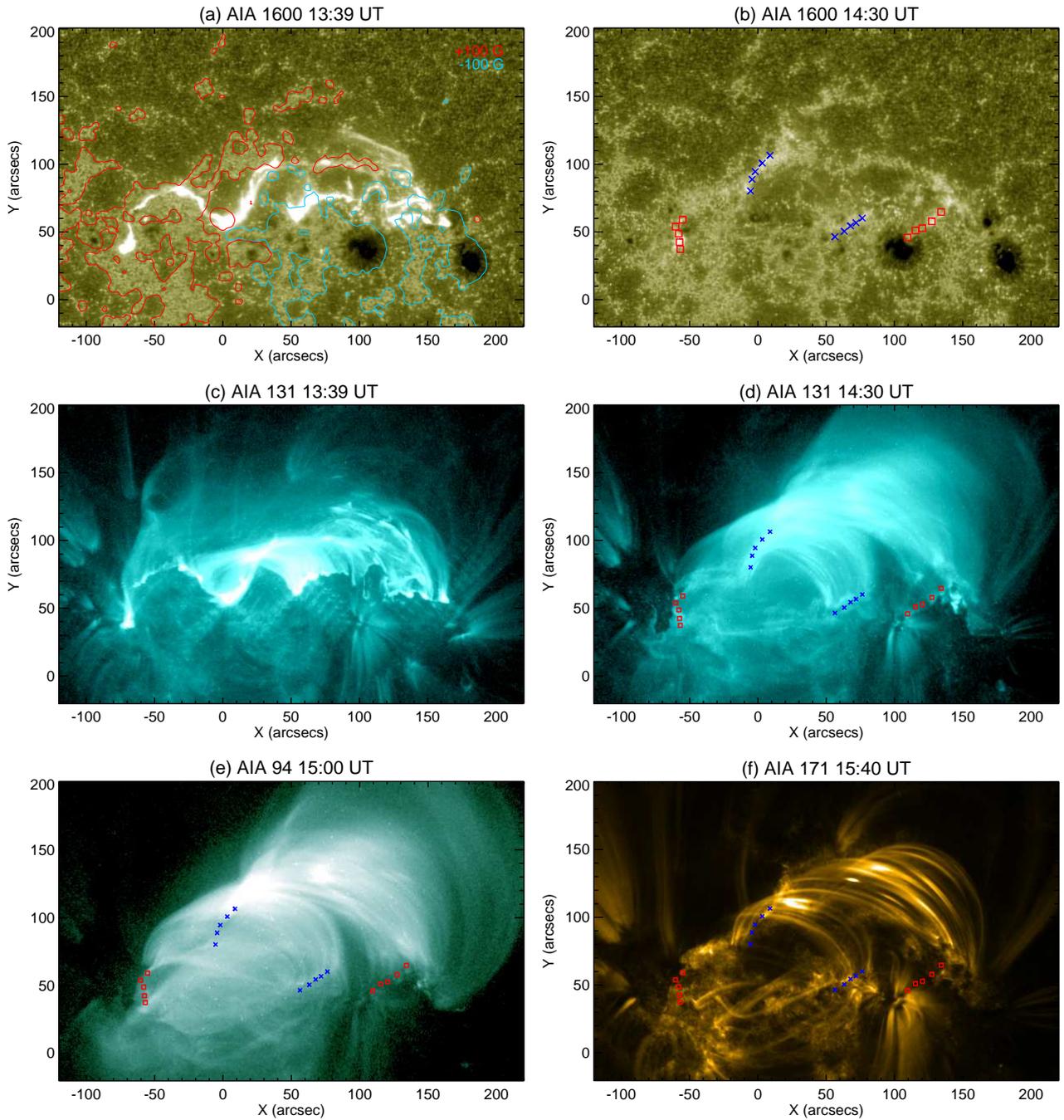}
	\caption{Connectivity of flare loops. Panels (a \& b) show flare ribbons in 1600~{\AA} during the rope-buildup and rope-erosion stage, respectively. (a) is superimposed by contours of a line-of-sight magnetogram at the levels of $\pm$100~G in red (positive polarity) and cyan (negative polarity). In (b) straight ribbons are marked by blue crosses and hooked ribbons by red squares. These symbols are replotted in (d--f). Panel (c) shows the 131~{\AA} image taken at the same time as (a). Panels (d--f) show an AIA 131, 94, and 171~{\AA} image, respectively, taken at successively later times. Post-flare loops of cooler temperatures dominate at later times, but their connectivities with respect to the hooked and straight flare ribbons are consistent with each other.  \label{fig:connectivity}}
\end{figure*}

\begin{figure*}
	\centering
	\includegraphics[width=\textwidth]{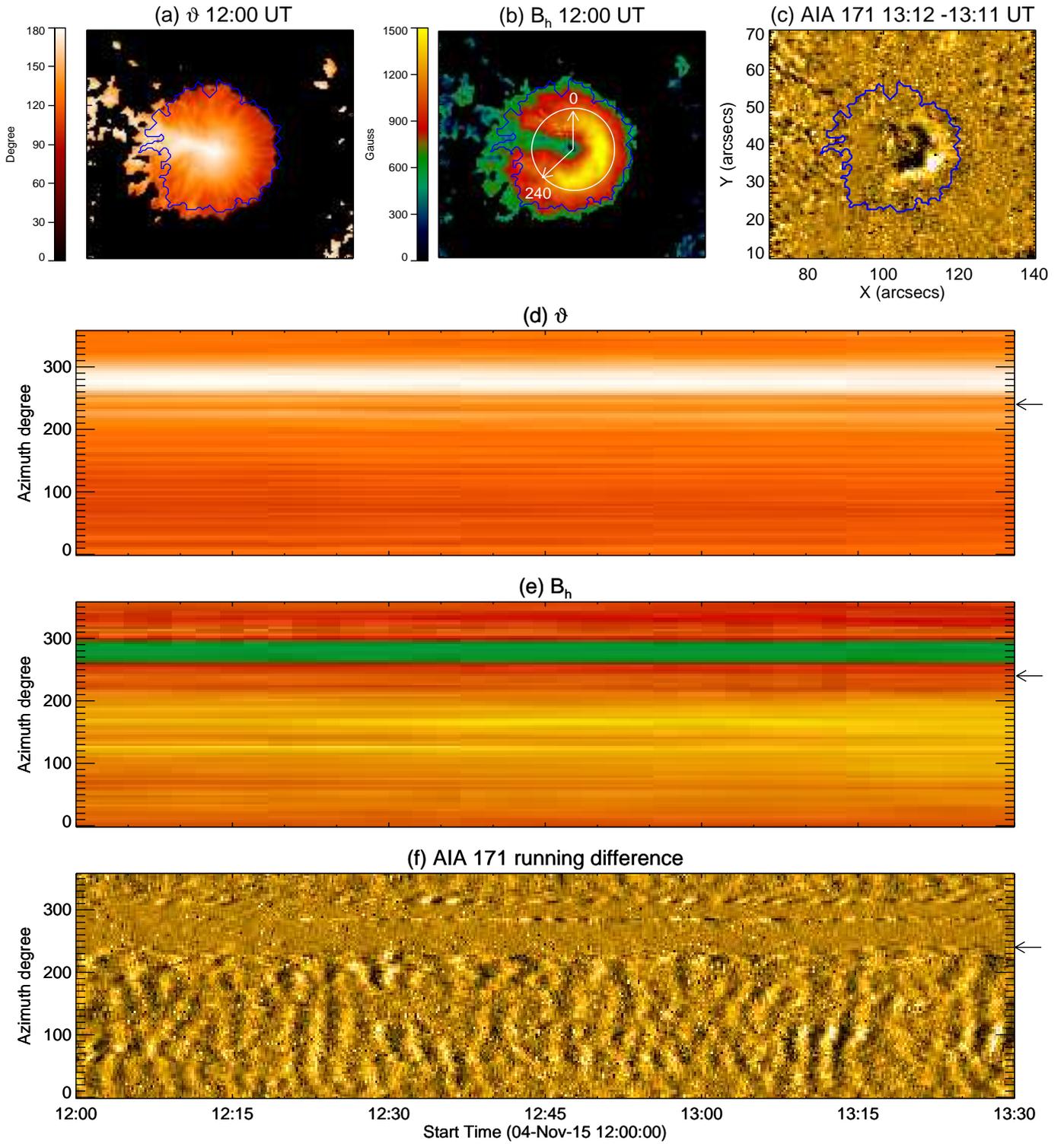}
	\caption{Pre-eruption RPWs in relation to magnetic field in the sunspot. Panels (a \& b) show field inclination $\theta$ and horizontal component $B_h$ of the photospheric magnetic field. Panel (c) shows the RPW front in an AIA 171~{\AA} difference image. The blue contour marks the outer penumbral boundary. The circle in (b) indicates the circular slit used to construct the time-distance maps in (d--f). The azimuthal directions of 0 and 240 deg along the circle are marked by two arrows. The latter direction is also marked by arrows on the right side in (d--f) .    \label{fig:wave-quiet}} 	
\end{figure*}

\begin{figure*}
	\centering
    \includegraphics[width=\textwidth]{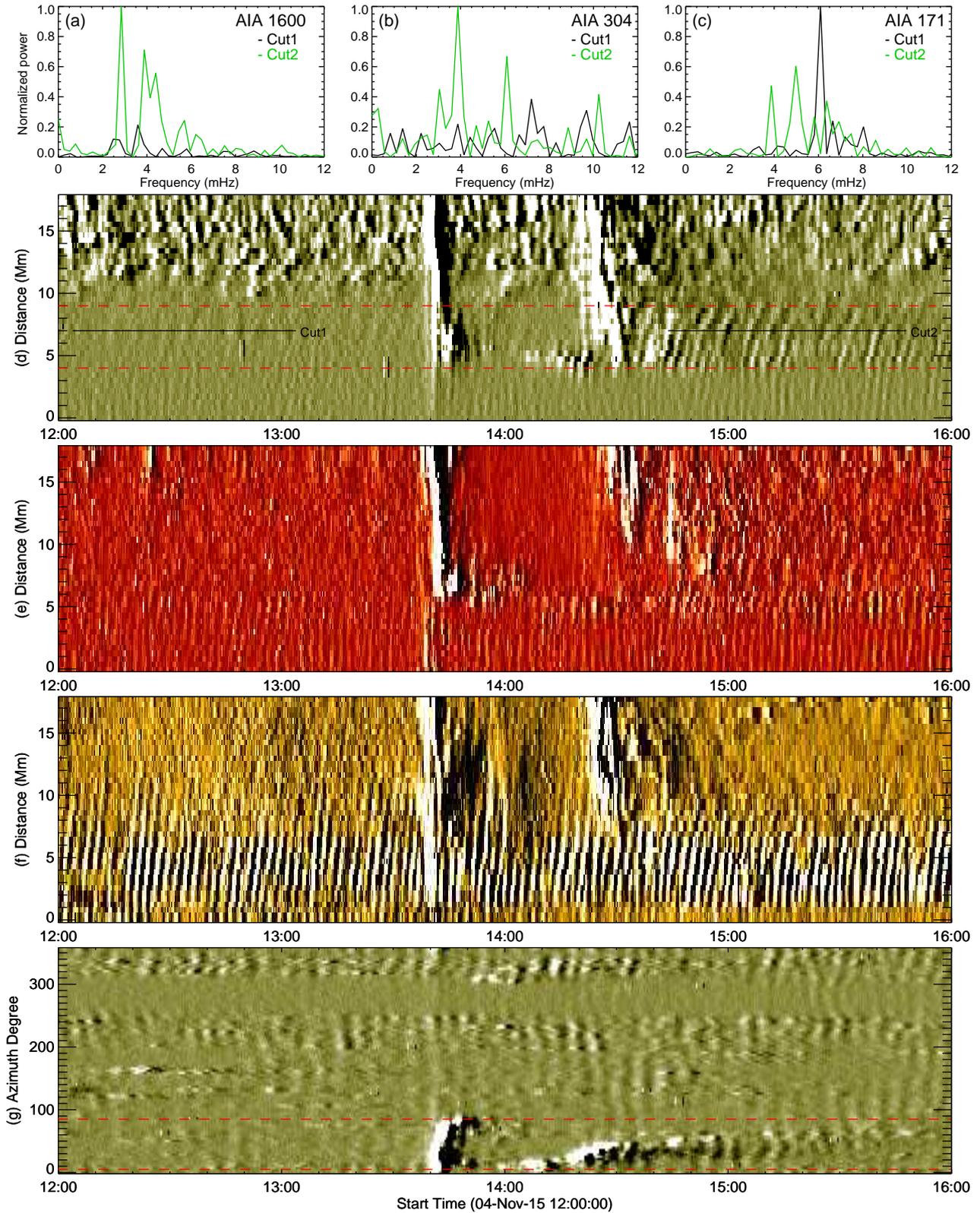}
    \caption{\label{fig:wave-erupt} RPWs observed across the MFR eruption. Panels (a--c) show power spectra of time series along Cut1 (black) and Cut2 (green) in the time-distance maps in panels (d-f), which are constructed by applying the linear slit (Figure~\ref{fig:fp-evolution}d) to 1600, 304, and 171~{\AA} images, respectively. The time-distance map in (g) is made by applying the circular slit (Figure~\ref{fig:wave-quiet}b) to 1600~{\AA} images. The PROI is bounded by two red dashed lines in (d) and (g). } 
\end{figure*}

We calculated the FFT power of time series obtained from horizontal cuts before the eruption (Cut1) and after the 2nd episode of ribbon sweeping (Cut2) in the time-distance maps at 7~Mm from the sunspot center. The short time interval between the two episodes of ribbon sweeping is not considered for FFT analysis, because the contamination of diffraction patterns and interference from eruption in 171~{\AA}. The FFT power of the pre-eruption RPWs in 171~{\AA} peaks at 6~mHz, but that of the post-eruption RPWs shifts to lower frequencies, with two significant peaks at 4 and 5~mHZ, respectively (Figure~\ref{fig:wave-erupt}c). Before the eruption, RPWs are not quite visible in the chromosphere, but they appear exclusively in the PROI in both 1600 and 304~{\AA} after being swept by HR$_\mathrm{E}$ at about 14:30 UT (Figure~\ref{fig:wave-erupt}(d \& e); see also Figure~\ref{fig:ribbon-morphology}(c \& d)). Their pre-eruption FFT power spectra are both noisy. But after the PROI is swept by HR$_\mathrm{E}$, the 1600~{\AA} power spectrum displays a primary peak at 3~mHz and a secondary peak at 4~mHZ, while the 304~{\AA} power spectrum has a primary peak at 4~mHz and a secondary peak at 6~mHz (Figure~\ref{fig:wave-erupt}(a \& b)). 

We also applied the same circular slit as in Figure~\ref{fig:wave-quiet} to 1600 running difference images. In the resultant time-distance maps (Figure~\ref{fig:wave-erupt}g), the sweeping of HR$_\mathrm{S}$ at about 13:40 UT can be seen as a brightening across the PROI as bounded by red dashed lines, and the sweeping of HR$_\mathrm{E}$ is shown as a slanted white stripe between 14:00--14:30 UT. Appearing as alternating black-and-white short stripes in running difference images, the RPW fronts are observed in enhanced contrast within the PROI, but not in other regions (see also Figure~\ref{fig:ribbon-morphology}c), right after the 2nd episode of ribbon sweeping. This enhancement of RPWs in the PROI is best observed in 1600~{\AA}, but also marginally visible in other passbands featuring the chromosphere (304~{\AA}; Figure~\ref{fig:ribbon-morphology}d) and upper photosphere (1700~{\AA}, not shown).

\subsection{Evolution of phostospheric magnetic field in the PROI} \label{subsec:field_change}
To investigate the field changes in relation to the flaring process, the two episodes of hooked ribbon sweeping the PROI are indicated by the enhanced 1600~{\AA} brightness averaged over the PROI (Figure~\ref{fig:bproi}a), same as the blue curve in Figure~\ref{fig:fp-evolution}a. The average $|B_r|$ and $B_h$ in the PROI both increase simultaneously with HR$_\mathrm{S}$ approaching the PROI at about 13:40 UT; the increasing rate is maintained at about 30 and 40 G~hr$^{-1}$ for $|B_r|$ and $B_h$, respectively, lasting about two hours. Meanwhile, both direct (DC) and return (RC) current in the PROI increase transiently during the 1st episode, when HR$_\mathrm{S}$ sweeps through the PROI; but not during the 2nd episode, when HR$_\mathrm{E}$ sweeps through the PROI. In this case, DC has the same negative sign as $B_r$ because the flux rope possesses positive helicity, as verified by the in-situ investigation of the magnetic cloud (Paper I). Since the expansion of FP- lasts only about 10 min (Figure~\ref{fig:fp-evolution}), the associated change in magnetic fields as well as electric currents could be missed if the conventional 720-s vector magnetograms were used. 
	
The detailed field evolution is more complicated than the overall trend revealed by Figure~\ref{fig:bproi}. Figure~\ref{fig:deltab}(a1--a3 \& b1--b3) show field changes relative to 13:52 UT, the watershed between the rope-buildup and erosion stage (marked by the vertical dashed line), in field inclination $\theta$ and radial ($|B_r|$) and horizontal ($B_h$) components, respectively. The 1st episode of ribbon sweeping  (13:40--13:45 UT) during the rope-buildup stage does not leave lasting imprint on the photospheric field in the PROI (Figure~\ref{fig:deltab}(a1--a3)). But during rope-erosion stage, one can see that $B_h$ is enhanced over the whole PROI (Figure~\ref{fig:deltab}(b3)). Meanwhile, $|B_r|$ increases at the penumbral filaments where RPWs are visible in the corona (Figure~\ref{fig:deltab}(b2)), see also Figure~\ref{fig:wave-quiet}c). Thus, the field inclination $\theta$ also tends to increase at these penumbral filaments (Figure~\ref{fig:deltab}(b1)).
	
Applying the same linear slit as in Figure~\ref{fig:fp-evolution}d to HMI vector magnetograms at 135-s cadence, we show changes of $\theta$, $|B_{r}|$, and $|B_{h}|$ across the PROI relative to before the eruption (13:00 UT) in time-distance maps (Figure~\ref{fig:deltab}(c--e)). The inner and outer boundary of the PROI along the slit are marked by two horizontal dashed lines. Again, one can see that along this particular slit, both $|B_{r}|$ and $|B_{h}|$ are enhanced exclusively within the PROI during the rope-erosion stage. The inclination $\theta$ is also increased by a few degrees along this slit, suggesting that the relevant field vectors turn slightly more vertical as $\theta$ is in the range  $(90^\circ$--$180^\circ)$. However, decreases in $\theta$ can be found in other places in the PROI (Figure~\ref{fig:deltab}(b1)). 

The temporal evolution of electric current and magnetic flux within the flux-rope footpoint areas in the same event has been investigated in Paper I (see Supplementary Figure 4 therein). Both electric current and magnetic flux decrease with time during the rope-erosion stage, which is consistent with similar studies such as \citet{Cheng&Ding2016,Barczynski2020,Xing2020}.

\section{Discussion \& Conclusion} \label{sec:dis}

To summarize, we investigated the changes of RPWs during the two-stage evolution of an erupting MFR, associated with two episodes of ribbon sweeping through a section of the sunspot penumbra (denoted as PROI). The rope's feet are outlined by two closed, hooked ribbons at far ends of the two straight flare ribbons. We focused on the rope's western foot of negative polarity, FP-, which temporarily overlaps the sunspot at the PROI. During the rope-buildup stage, FP- expands and the southern section of the hooked ribbon rapidly sweeps through the PROI. Thus, the original penumbral field in the PROI is converted into the rope field, which is associated with a transient increase in electric currents flowing through the PROI. During the rope-erosion stage, FP- shrinks and the eastern section of the hooked ribbon slowly sweeps through the PROI, which converts the rope field in the PROI to post-flare loops and is associated with a gradual increase in $B_h$ in the PROI. Meanwhile, post-eruption RPWs display extra low-frequency components in the 3--5~mHz range (\S\ref{subsec:post-erupt-rpw}), compared with the pre-eruption counterpart that peaks at 6~mHz. Such changes at the footpoint area of an erupting flux rope have not been reported before. The interpretation of these changes and their implications are discussed below.

\subsection{Changes of RPWs in response to ribbon sweeping}
After the 2nd episode of ribbon sweeping, the RPWs in the PROI extend their `bandwidth' to lower frequencies down to 3~mHz, which can be understood if the cutoff frequency $f_c$ is reduced in the PROI. However, the effect of the varying magnetic field on $f_c$ is mixed in the PROI: the field of penumbral filaments turns more vertical, while the horizontal field component is overall enhanced. Consequently $\cos\theta$ increases at penumbral filaments but decreases at intra-filament regions. Thus, the reduction of $f_c$ must be mainly caused by an increase in plasma temperature. This is evidenced by the diffuse 1600~{\AA} emission lingering in the PROI after the ribbon sweeping (Figure~\ref{fig:fp-evolution}a). This diffuse emission is most likely contributed by the C~\rom{4} doublet (1550~{\AA}; $\log T =5$) in the 1600~{\AA} passband, which is excited as the reconnection that converts the rope field to post-flare arcade field heats up the arcade footpoints in the PROI to transition-region temperatures. The excitation of the C~\rom{4} doublet may also account for the enhanced contrast of RPW fronts observed in 1600~{\AA}, because these fronts are visible in the best contrast in the upper transition region as featured by the AIA 171~{\AA} passband. 

\subsection{Changes of magnetic field in response to ribbon sweeping}
The evolution of penumbral field in response to the sweeping of the hooked ribbon is subtle, and the ribbon-swept penumbra does not change its appearance. In contrast, e.g., \citet{WangH2012} found in H$\alpha$ observations that in a ribbon-swept penumbral region, dark fibrils disappear and bright grains evolve into faculae, which are interpreted as the penumbral field turning consistently from horizontal to vertical, leading to the decay of the ribbon-swept penumbra. These differences between our case and that in \citet{WangH2012} may be attributed to different physical processes occurring at different segments of flare ribbons. In our case it is the hooked ribbon, but in \citet{WangH2012} it is the straight ribbon (their Figure 2), that sweeps the penumbra. 
	
\begin{figure*}
	\centering
	\includegraphics[width=\textwidth]{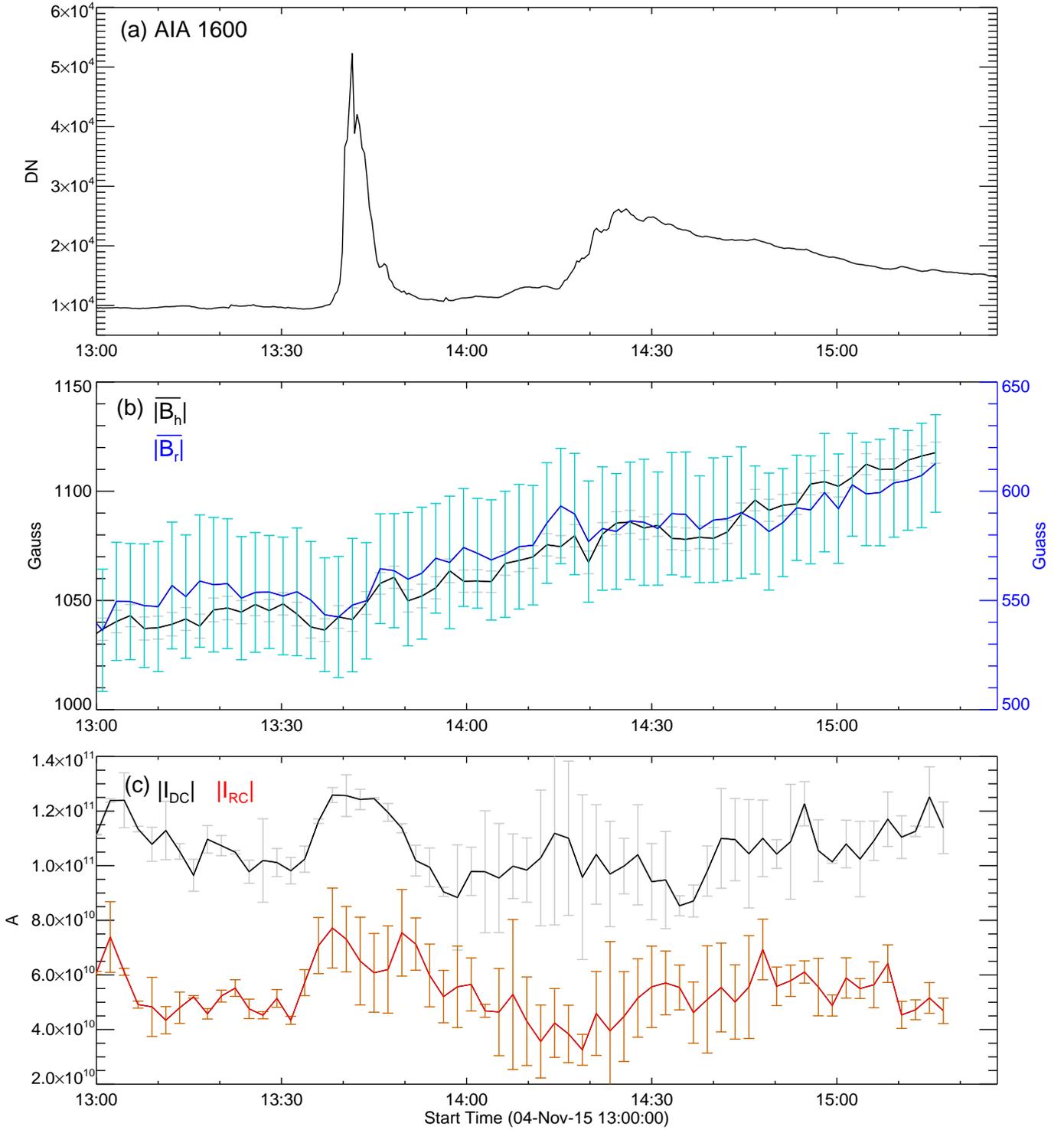}
	\caption{\label{fig:bproi} Evolution of photospheric magnetic field in the PROI in relation to the flare. a) Average brightness of the PROI in AIA 1600~{\AA}; b) average $|B_r|$ (scaled by the left $y$-axis) and $B_h$ (scaled by the right $y$-axis) in the PROI; c) direct ($|I_\text{DC}|$) and return ($|I_\text{RC}|$) current flowing through the PROI. } 
\end{figure*}

\begin{figure*}
	\centering
	\includegraphics[width=\textwidth]{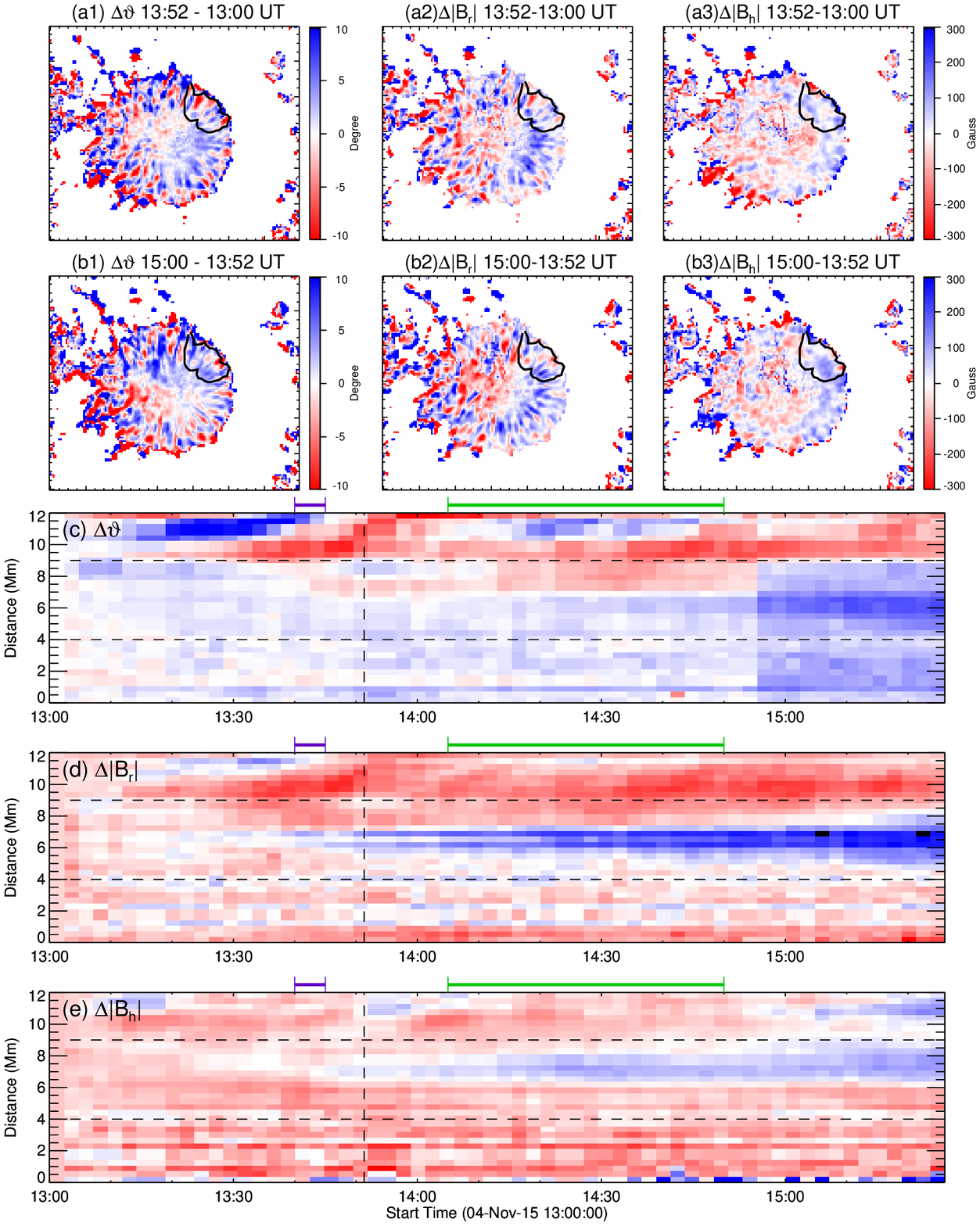}
	\caption{\label{fig:deltab} Changes of photospheric magnetic field across the MFR eruption. Panels (a1--a3) and (b1--b3) show the changes of inclination $\theta$, radial component $B_r$, and horizontal component $B_h$ of photospheric magnetic field after the 1st and 2nd episode of ribbon sweeping, respectively. The PROI is outlined by the closed curve. The time-distance maps in (c--e) are made by applying the linear slit (Figure~\ref{fig:fp-evolution}d) to HMI vector magnetograms subtracted by a reference magnetogram at 13:00 UT. The purple and green bars on the top of (c--e) indicates the two episodes of ribbon sweeping during the rope-buildup and rope-erosion stage, respectively. $\delta\theta$ is color-coded by the color bar of the same as in (a1) and (b1), while change } 
\end{figure*}

Associated with the 2nd episode of hooked ribbon sweeping the PROI during the flare decay phase, both radial and horizontal components of photospheric field increase in magnitude in a gradual fashion in the ribbon-swept region (Figure~\ref{fig:bproi}). These observations cannot be easily explained by the momentum conservation of CMEs nor by the implosion theory, both of which intend to address the sudden change in photospheric magnetic field around the flaring PIL --- between the two straight flare ribbons representing the footpoints of the post-flare arcade, where the horizontal field is enhanced stepwise across the flare impulsive phase \citep[e.g.,][]{WangS2012,Sun2012,liu2018evolution}. Instead, our observations can be understood by magnetic reconnections occurring at the hooked ribbon, which is the footpoint locus of miscellaneous field lines that may belong to either the flux rope or the post-flare arcade, as they are interchangeable with each other under the ongoing magnetic reconnection between the flux rope and the ambient field, which is manifested as the hooked ribbon sweeping back and forth across the PROI.

According to the terminology introduced by \cite{Aulanier2019drifiting}, magnetic reconnection involved in the 1st episode of ribbon sweeping can be denoted as `aa-rf', i.e., two arcade field lines turn into a rope field line and a flare loop, which results in the buildup of the MFR in the corona. Interestingly, this episode is associated with a transient increase in electric currents flowing through the PROI, consistent with the concept that a magnetic flux rope is a current-carrying structure.  Magnetic reconnection involved in the 2nd episode could be either `rr-rf' or `ar-rf', in which field lines of the MFR reconnect with each other or with the ambient arcade at the hooked ribbon into new MFR field lines and flare loops. In addition to heating up the low atmosphere at the footpoints of flare loops, these magnetic reconnections result in the erosion and/or drifting of the MFR feet while modifying the photospheric magnetic field. 
	
As far as field inclination is concerned, supposedly, a highly twisted field line becomes more vertical when it is transformed into an arcade field line by reconnection; the reverse transformation makes the field line more horizontal. But the transformation between a weakly twisted field line and an arcade field line may not produce a significant difference in the photosphere. However, if the twisted field line is already highly stretched due to the eruption, then it generally turns more horizontal when being transformed into an arcade field line. The situation is thence further complicated by the rope's nonuniform twist distribution, which is demonstrated to be highly related to the reconnection rate during the rope-buildup stage \citep{wang2017buildup}. Thus, the change in field inclination can be modulated by both the magnetic twist and stretched state, i.e., the local twist density, of flux-rope field lines, by whether the same field lines are turned into, or transformed from, arcade field lines by reconnection, and by the relaxation process of the newly reconnected field lines. These complexities may be manifested as the complex field changes observed in the ribbon-swept region \citep[\S\ref{subsec:field_change}; see also][]{liu2018evolution}. 

Overall, since it is not a rare occasion for flare ribbons to move into sunspots, the resultant changes in sunspot waves and oscillations as well as in photospheric field may offer a new window to diagnose flare reconnections.


\begin{acknowledgements}
	This work was supported by the National Natural Science Foundation of China (NSFC; Grant Nos. 41761134088, 41774150, and 11925302) and the Strategic Priority Program of the Chinese Academy of Sciences (Grant No. XDB41030100). W.W. is supported by the China Postdoctoral Science Foundation (Grant No. 2019TQ0313). 
\end{acknowledgements}

\end{document}